%% file: covariant_LQBH_paperversionArxiV3.tex
\documentclass[aps,prd,reprint,superscriptaddress,showpacs,footinbib]{revtex4-1}

\usepackage{amssymb,amsmath,amssymb,amsfonts,amsthm,stmaryrd,mathrsfs,physics,bbm,mfirstuc}
\usepackage{graphicx}
\usepackage[T1]{fontenc}
\usepackage{enumerate} % advanced enumerate environment
\usepackage{color} % for colored text

\usepackage{hyperref} % for HyperTeX cross-referencing

\newcommand{\heff}{{H_{\rm eff}}}

\begin{document}

\title{Black Holes and Covariance in Effective Quantum Gravity
%Black Holes and Covariance of Semiclassical Models
}

\author{Cong Zhang}
%\email{zhang.cong@mail.bnu.edu.cn}
\affiliation{School of Physics and Astronomy, Key Laboratory of Multiscale Spin Physics,
Ministry of Education, Beijing Normal University, Beijing 100875, China}
\affiliation{Department Physik, Institut f\"ur Quantengravitation, Theoretische Physik III, Friedrich-Alexander-Universit\"at Erlangen-Nürnberg, Staudtstra{\ss}e 7/B2, 91058 Erlangen, Germany}

\author{Jerzy Lewandowski}
%\email{jerzy.lewandowski@fuw.edu.pl}
\affiliation{Faculty of Physics, University of Warsaw, Pasteura 5, 02-093 Warsaw, Poland}

\author{Yongge  Ma}
\email{mayg@bnu.edu.cn}
\affiliation{School of Physics and Astronomy, Key Laboratory of Multiscale Spin Physics,
Ministry of Education, Beijing Normal University, Beijing 100875, China}

\author{Jinsong Yang}
\email{jsyang@gzu.edu.cn}
\affiliation{School of Physics, Guizhou University, Guiyang 550025, China}

\begin{abstract}
The longstanding issue of general covariance in effective models of quantum gravity is addressed, which arises when canonical quantum gravity leads to a semiclassical model described by an effective Hamiltonian constraint. In the context of spherically symmetric models, general covariance is precisely formulated into a set of equations, leading to the necessary and sufficient conditions for ensuring covariance. With the aid of these conditions, we derive the equations for the effective Hamiltonian constraint. The equations yield two candidates for effective Hamiltonian constraints dependent on a quantum parameter. The resulting quantum modified black hole spacetimes are analyzed. Our models show improvement by casting off the known limitations of previous works with similar results.
\end{abstract}

%\keywords{Black hole, loop quantum gravity, singularity resolution }

\maketitle
%\tableofcontents

\section{Introduction}
Despite its successes, general relativity (GR) faces challenges, including the existence of singularities \cite{PhysRevLett.14.57} and its incompatibility with quantum physics \cite{polchinski1998string,Ambjorn:2001cv,ashtekar2015general,Surya:2019ndm}, suggesting it may not be the final theory of spacetime. This prompts ongoing efforts to modify GR. Beyond existing approaches \cite{Clifton:2011jh,Mukohyama:2019unx,Yao:2020tur}, we explore modifications within the Hamiltonian formulation while maintaining 4-dimensional diffeomorphism covariance, which is particularly relevant for semiclassical gravity emerging from canonical quantum gravity models.

In GR, diffeomorphism covariance, when passing to the Hamiltonian formulation, is encoded in the Poisson algebra of the constraints. The reverse issue, i.e., the \textit{covariance issue},  arises: under what conditions does a given $3+1$ model in the Hamiltonian formulation describe a generally covariant spacetime theory. The covariance, initially investigated in, e.g., \cite{Bojowald:2011aa}, is a general issue in any effective Hamiltonian theory resulting from a canonical quantum theory of gravity. For instance, this issue has been long debated \cite{Tibrewala:2013kba,Bojowald:2015zha,Wu:2018mhg,Bojowald:2019dry,Bojowald:2020unm,Han:2022rsx,Gambini:2022dec,Bojowald:2022zog,Ashtekar:2023cod,Giesel:2023hys,Bojowald:2024beb} in the effective theories resulting from symmetry-reduced models of loop quantum gravity (LQG) \cite{Ashtekar:2006rx,Gambini:2008dy,Gambini:2013hna,Ashtekar:2018lag,Zhang:2020qxw,Husain:2021ojz,Alonso-Bardaji:2021yls,Zhang:2021wex,Han:2022rsx,Gambini:2022hxr,Husain:2022gwp,Alonso-Bardaji:2022ear,Giesel:2023tsj,ElizagaNavascues:2023gow,Bojowald:2023djr,Bojowald:2023vvo,Li:2023axl,Alonso-Bardaji:2023vtl}.

In the previous works \cite{Zhang:2021wex, Zhang:2021xoa, Husain:2021ojz, Han:2022rsx,Husain:2022gwp,Giesel:2023hys,Giesel:2023tsj}, a specific matter field is chosen to fix the gauge of diffeomorphism.  Then, quantizing those models results in effective Hamiltonians valid within the preferred gauge.
In contrast, we aim to relax the reliance on the gauge-fixing with matter fields, and seek for effective Hamiltonian constraints valid uniformly across different gauges. Specifically, we focus on spherically symmetric gravity. While the conditions of general covariance are derived in vacuum case, the coupling to various matter fields is also taken into account to ensure broad applicability across diverse models. The solutions for these conditions are also obtained. The resulting spacetime metrics of these models are then analyzed.

\section{Approach to the covariance issue}
The kinematical structure which we assume is the same as that of the spherically symmetric GR on a 4-dimensional manifold $\mathcal M_2\times \mathbb S^2$, with $\mathbb S^2$ denoting the 2-sphere. By this symmetry, the theory is reduced to dilaton gravity on the 2-manifold  $\mathcal M_2\cong \mathbb R\times\Sigma$ with $\Sigma$ being the spatial manifold, which, for example, could be the positive real line \cite{Grumiller:2002nm}. 
Let $(t,x)$ denote the coordinate of $\mathcal M_2$ adapted to the topology. With this coordinate, $\mathcal M_2$ is foliated into constant-$t$ slices $\Sigma_t$, each diffeomorphic to $\Sigma$.
As shown in, e.g., \cite{Bojowald:2004af,Bojowald:2005cb,Gambini:2013hna,Zhang:2021xoa}, the phase space of the model contains the canonical pairs $(K_2,E^2)$ for the 2-dimensional gravity and $(K_1,E^1)$ for the dilaton. Note that the fields $E^2$ and $K_1$ are scalar densities with weight $1$ on $\Sigma$, while $K_2$ and $E^1$ are scalars on  $\Sigma$.  The Poisson brackets of the canonical pairs read $\{K_1(x),E^1(y)\}=2\delta(x,y)$ and $\{K_2(x),E^2(y)\}=\delta(x,y)$, where the geometrized units with $G=1=c$ are applied.

The dynamics is encoded in the diffeomorphism constraint $H_x$ and the Hamiltonian constraint $\heff$. Here $H_x$ is assumed to retain its classical form $H_x=E^2\partial_xK_2-K_1\partial_xE^1/2,$
which  generates the diffeomorphism transformations on $\Sigma$. However, $\heff$ is expected to deviate from the classical one due to some quantum gravity effects. Therefore, we will consider it unknown that needs to be determined. These constraints are assumed to be first class, 
and the constraint algebra is expected to mirror the classical ones with a correction factor $\mu$ to account for quantum gravity effects:
\begin{subequations}\label{eq:constraintalgebra}
\begin{align}
\{H_x[N^x_1],H_x[N^x_2]\}&=H_x[N^x_1\partial_xN^x_2-N^x_2\partial_xN^x_1]\label{eq:HxHx},\\
\{H_x[N^x_1],\heff[N_1]\}&=\heff[N^x_1\partial_xN_1]\label{eq:HxH},\\
\{\heff[N_1],\heff[N_2]\}&=H_x[S(N_1\partial_xN_2-N_2\partial_xN_1)]\label{eq:HH},
\end{align}
\end{subequations}
with $S\equiv \mu E^1(E^2)^{-2}$ being the structure function. Here, $N_1$ and $N_2$ are arbitrary smearing functions, and $N_1^x$ and $N_2^x$ are smearing vector fields. The notation $F[g] \equiv \int F(x) g(x) \dd x$ is used for convenience. This form of modified algebra \eqref{eq:constraintalgebra} has been considered in, e.g., \cite{Bojowald:2011aa,Giesel:2023tsj,AlonsoBardaji:2023bww}.

Choosing a specific lapse function $N$ and shift vector $N^x$, for given $\heff$, one can solve the Hamilton's equations using $\heff[N]+H_x[N^x]$ as the Hamiltonian, with some 
 initial data satisfying the constraints $\heff(x)=0=H_x(x)$. This yields a family of fields $(K_I^{(t)}(x),E^I_{(t)}(x))$ on $\Sigma$, parametrized by $t\in \mathbb R$. By  pushing forward the fields  for each $t$ from $\Sigma$ to $\Sigma_t$, one obtains $K_I(t,x)\equiv K_I^{(t)}(x)$ and $E^I(t,x)\equiv E^I_{(t)}(x)$,  as well as $N(t,x)\equiv N^{(t)}(x)$ and $N^x(t,x)\equiv N^x_{(t)}(x)$, which are now fields on $\mathcal M_2$. Here, $N$ and $N^x$ can  vary with $t$ because they can depend on $K_I$ and $E^I$.  Then one faces the covariance issue: Is it possible to construct a non-degenerate and symmetric tensor $g_{\rho\sigma}$ on $\mathcal M_2\times \mathbb S^2$ from the fields $N(t,x)$, $N^x(t,x)$, $E^I(t,x)$ and $K_I(t,x)$ in such a manner that $g_{\rho\sigma}$ remains invariant  up to diffeomorphisms of  $\mathcal M_2$ for different choices of $N$ and $N^x$? If this condition is satisfied for some $g_{\rho\sigma}$, the theory is called being covariant with respect to this $g_{\rho\sigma}$. Here, only the diffeomorphisms of  $\mathcal M_2$ are considered because they preserve the spherical symmetry.

\subsection{Sufficient and necessary conditions for covariance}
When $\heff$ is the classical Hamiltonian constraint, the theory will be covariant with respect to the classical metric $g_{\rho\sigma}$. However, for a general $\heff$, the associated theory may not be covariant with respect to $g_{\rho\sigma}$ any more. Thus, we explore the possibility of modifying $g_{\rho\sigma}$ to define an effective metric $g_{\rho\sigma}^{(\mu)}$, with respect to which the theory becomes covariant.

A reasonable modification arises from analyzing the geometric meaning of the constraint algebra \eqref{eq:constraintalgebra} as shown in, e.g., \cite{Bojowald:2011aa}. In the classical theory where $\mu=1$, the structure function $S$, becoming $E^1(E^2)^{-2}$, is the $(x,x)$-component of the inverse spatial  metric of $g_{\rho\sigma}$, so that the constraint algebra describes the behavior of hypersurface deformations. If we assume that the new $S$ in  \eqref{eq:HH} represents the $(x, x)$-component of the inverse spatial metric of $g_{\rho\sigma}^{(\mu)}$, the algebra \eqref{eq:constraintalgebra}  retains the same geometric interpretation as in the classical case. This motivates us to define $g_{\rho\sigma}^{(\mu)}$ by
\begin{equation}\label{eq:g4eff}
\dd s^2=- N^2\dd t^2+\frac{(E^2)^2}{\mu E^1}(\dd x+N^x\dd t)^2+E^1\dd\Omega^2,
\end{equation}
where $\dd\Omega^2=\dd\theta^2+\sin^2\theta\dd\phi^2$  is the fiducial metric on $\mathbb S^2$.

To ensure covariance with respect to $g_{\rho\sigma}^{(\mu)}$, we seek the effective Hamiltonian constraint $\heff$ such that 
\begin{equation}\label{eq:covariancedef}
\delta g_{\rho\sigma}^{(\mu)} = \mathcal{L}_{\alpha \mathfrak{N}} g_{\rho\sigma}^{(\mu)},
\end{equation}
up to terms proportional to constraints. Here, $\mathfrak{N} = \partial_t - N^x \partial_x$, and $\delta g_{\rho\sigma}^{(\mu)}$ represents the infinitesimal gauge transformation generated by $\heff[\alpha N]$, where $\alpha$ and $N$ could be phase space dependent.
A straightforward calculation (see Appendix \ref{app:A}) shows that Eq.~\eqref{eq:covariancedef} is equivalent to the following two conditions: (i) $\heff$ is independent of derivatives of $K_1$; and (ii) $\{S(x), \heff[\alpha N]\}=\alpha(x)\{S(x),\heff[N]\}$ for any phase space independent  functions $\alpha$ and $N$. These conditions are derived by using the constraint algebra \eqref{eq:constraintalgebra} and the fact that $H_x$ generates the spatial diffeomorphism transformation. Therefore, when matter is coupled, the conditions remain applicable but with $\heff$ replaced by the total effective Hamiltonian constraint.

\subsection{Covariance equations}
Since $\heff$ is a scalar density of weight $1$, akin to $E^2$, it can generally be expressed as 
\begin{equation}
\heff=E^2 F,
\end{equation}
with some scalar $F$.  To construct $F$, we consider basic scalars formed from the phase space variables $K_I$, $E^I$, and their derivatives. 
Due to the condition (i), the derivatives of $K_1$ will be excluded. Similarly, it is natural to omit the derivatives of $K_2$ as $K_1$ and $K_2$ play analogous roles in the model. 
This ansatz aligns with the cases in classical GR and several widely studied quantum black hole (BH) models (see, e.g., \cite{Alonso-Bardaji:2021yls,Han:2022rsx,Husain:2022gwp,Giesel:2023hys,Bojowald:2023djr}).
 Furthermore, derivatives of $E^I$ higher than the second order are not allowed,  since Eq.~\eqref{eq:HH} is expected to hold  with the right-hand side proportional to $N_1\partial_xN_2-N_2\partial_xN_1$ while $K_I$ must be involved in $\heff$. Suppose that $\partial_x^nE^I$ is included in $F$. Then the result of $\{\heff[N_1],\heff[N_2]\}$ will contain terms of the form $\int\dd x f(N_1\partial_x^nN_2-N_2\partial_x^nN_1)$ with $f$ depending on $K_I$ and $E^I$. These terms are proportional to $(N_1\partial_xN_2-N_2\partial_xN_1)$ for $n\leq 2$, but not for $n>2$. For $n=2$, the integration by parts can be used to simplify the expression to $\int \dd x(\partial_xf)(N_1\partial_xN_2-N_2\partial_xN_1)$. With these considerations, the only remaining fundamental scalars are as follows:
\begin{equation*}
\begin{aligned}
&s_1=E^1,\ s_2=K_2,\ s_3=\frac{K_1}{E^2},\ s_4=\frac{\partial_xs_1}{E^2},\\
&s_5=\frac{\partial_xs_4}{E^2},\ s_6=\frac{\partial_xs_1}{K_1},\ s_7=\frac{\partial_xs_4}{K_1}.
\end{aligned}
\end{equation*} 
Additionally, to ensure that the effective model accommodates solutions analogous to the Schwarzschild solution in classical GR, which allows the $3+1$ decompositions such that either the corresponding $K_I$ or the derivatives of $E^I$ vanish across the entire $t$-slice, we exclude the basic scalars $s_6$ and $s_7$, since they are ill-defined for such  $3+1$ decompositions.

With $F$ depending on $s_a$ for $1\leq a\leq 5$, the Poisson bracket between $\heff[N_1]$ and $\heff[N_2]$ can be computed straightforwardly as 
\begin{equation}\label{eq:poissHH12}
\begin{aligned}
&\{\heff[N_1],\heff[N_2]\}=2\int D(N_1,N_2)\Big[\frac{F_2F_5 s_4}{2}\\
&-F_3F_4+\sum_{a=1}^5\left( F_3\partial_{s_a}F_5-F_5\partial_{s_a}F_3 \right)\frac{\partial_xs_a}{E^2}
  \Big],
\end{aligned}
\end{equation}
where we used the abbreviations $F_a\equiv \partial_{s_a}F$ and $D(N_1,N_2)\equiv N_1\partial_xN_2-N_2\partial_xN_1$. Substituting Eq.~\eqref{eq:poissHH12} into Eq.~ \eqref{eq:HH}, we obtain a differential equation for $F$, which can be solved by means of the two conditions aforementioned. 
As a result, $\heff$ is determined to take the form
\begin{equation}\label{eq:HeffOnMeff}
\heff=-2E^2\Big[\partial_{s_1}M_{\rm eff}+\frac{\partial_{s_2}M_{\rm eff}}{2}s_3+\frac{\partial_{s_4}M_{\rm eff}}{s_4}s_5+\mathcal R\Big],
\end{equation}
where $\mathcal R$ is an arbitrary function of $s_1$ and $M_{\rm eff}$, and $M_{\rm eff}$ depending on $s_1, s_2, s_4$ is a solution to:
\begin{subequations}\label{eq:covarianceMeff}
\begin{align}
&\frac{\mu  s_1 s_4}{4}=(\partial_{s_2}M_{\rm eff})\partial_{s_2}\partial_{s_4}M_{\rm eff}-(\partial_{s_4}M_{\rm eff})\partial_{s_2}^2M_{\rm eff},\label{eq:covarianceequationgeneral1}\\
&(\partial_{s_2}\mu)\partial_{s_4}M_{\rm eff}-(\partial_{s_2}M_{\rm eff})\partial_{s_4}\mu=0.\label{eq:covarianceequationgeneral2}
\end{align}
\end{subequations}
Note that Eqs.~\eqref{eq:covarianceequationgeneral1} and \eqref{eq:covarianceequationgeneral2} originate from the constraint algebra \eqref{eq:HH} and the covariance condition (ii), respectively.
According to Eq.~\eqref{eq:covarianceequationgeneral2}, $\mu$ depends only on $s_1$ and $M_{\rm eff}$. Note that, in the vacuum case, by requiring $\mathcal R=0$,  $M_{\rm eff}$ becomes a Dirac observable representing the BH mass (see \cite{note2,Giesel:2008zz} for Dirac observables). One can also choose $\mathcal R\propto \sqrt{s_1}$ to incorporate the contribution of the cosmological constant.

\section{Effective Hamiltonian constraints}
By solving the covariance equations \eqref{eq:HeffOnMeff} and \eqref{eq:covarianceMeff}, various solutions can be obtained, each corresponding to a covariant model. For example, the classical theory emerges as one particular solution to these equations. 

In the following discussion, we focus on solutions related to loop quantum BH models \cite{Husain:2021ojz,Husain:2022gwp,Giesel:2023hys,Giesel:2023tsj}, which are characterized by the polymerization procedure that replaces connections with their holonomies. This replacement reflects the loop representation, ensuring background independence in the resulting quantum theory \cite{Ashtekar:1995zh}.  In the current model, due to its symmetries, the holonomies can be simplified to combinations of trigonometric functions of $K_I$ \cite{Ashtekar:2006wn}. A particular polymerization is to substitute $s_2=K_2$ with $\zeta^{-1}\sqrt{s_1}\sin\left(\frac{\zeta s_2}{\sqrt{s_1}}\right)$, where $\zeta$, proportional to the Planck length $\sqrt{\hbar}$, is a quantum parameter.
Now we propose the following two effective masses involving the polymerization, leading to two different covariant models.

 The first effective mass reads
\begin{equation*}
M_{\rm eff}^{(1)}=\frac{\sqrt{s_1}}{2}+\frac{\sqrt{s_1}^3\sin^2\left(\frac{\zeta s_2}{\sqrt{s_1}}\right)}{2\zeta^2}-\frac{\sqrt{s_1}(s_4)^2}{8}e^{\frac{ 2 i \zeta s_2}{\sqrt{s_1}}},
\end{equation*}
which is a solution to Eq.~\eqref{eq:covarianceMeff} with $\mu\equiv \mu_1=1$. By substituting $M_{\rm eff}^{(1)}$ for $M_{\rm eff}$ in Eq.~\eqref{eq:HeffOnMeff}, we obtain the first effective Hamiltonian  constraint by setting $\mathcal R=0$:
\begin{equation}\label{eq:heff1}
\begin{aligned}
&H_{\rm eff}^{(1)}=-\frac{E^2}{2  \sqrt{E^1}}-\frac{K_1E^1}{2\zeta  } \sin \left(\frac{2\zeta  K_2}{\sqrt{E^1}}\right)\\
&-\frac{3 \sqrt{E^1}E^2}{2 \zeta ^2 } \sin ^2\left(\frac{\zeta  K_2}{\sqrt{E^1}}\right) +\frac{K_2E^2}{2 \zeta  } \sin \left(\frac{2\zeta  K_2}{\sqrt{E^1}}\right)\\
&+\frac{\left(\partial_xE^1\right)^2 }{8 \sqrt{E^1}E^2}e^{\frac{2 i \zeta  K_2}{\sqrt{E^1}}}+\frac{\sqrt{E^1}  }{2 }\partial_x\left(\frac{\partial_xE^1}{E^2}\right)e^{\frac{2 i \zeta  K_2}{\sqrt{E^1}}}\\
&+\frac{i \zeta  E^2}{4 }\left(\frac{\partial_xE^1}{E^2}\right)^2\left(\frac{K_1}{E^2}-\frac{K_2}{E^1}\right)e^{\frac{2 i \zeta  K_2}{\sqrt{E^1}}}.
\end{aligned}
\end{equation}
Notably, even though $H_{\rm eff}^{(1)}$ appears to be complex, as we will show below, the resulting effective metric is real provided that $K_I$ is allowed to take complex values to guarantee the reality of $M_{\rm eff}^{(1)}$.

The second effective mass is 
\begin{equation*}
M_{\rm eff}^{(2)}=\frac{\sqrt{s_1}}{2}+\frac{\sqrt{s_1}^3 \sin ^2\left(\frac{\zeta  s_2}{\sqrt{s_1}}\right)}{2\zeta ^2 }-\frac{\sqrt{s_1} (s_4)^2 \cos ^2\left(\frac{\zeta  s_2}{\sqrt{s_1}}\right)}{8},
\end{equation*}
which is the solution to Eq.~ \eqref{eq:covarianceMeff} with 
\begin{equation*}
\mu\equiv \mu_{2}=1+\frac{\zeta^2 }{\sqrt{s_1}^3}\left( \sqrt{s_1}-2M_{\rm eff}^{(2)}\right).
\end{equation*}
The corresponding effective Hamiltonian constraint with $\mathcal R=0$ reads
\begin{equation}\label{eq:heff2}
\begin{aligned}
&H_{\rm eff}^{(2)}=-\frac{E^2}{2  \sqrt{E^1}}-\frac{  K_1 E^1}{2\zeta  }\sin \left(\frac{2\zeta  K_2}{\sqrt{E^1}}\right)\\
&-\frac{3 \sqrt{E^1}E^2}{2 \zeta ^2 } \sin ^2\left(\frac{\zeta  K_2}{\sqrt{E^1}}\right)+\frac{K_2E^2 }{2\zeta  } \sin \left(\frac{2\zeta  K_2}{\sqrt{E^1}}\right)\\
&+\frac{\left(\partial_xE^1\right)^2}{8  \sqrt{E^1}E^2} \cos ^2\left(\frac{\zeta  K_2}{\sqrt{E^1}}\right)\\
&+\frac{\sqrt{E^1} }{2 }\partial_x\left(\frac{\partial_xE^1}{E^2}\right) \cos ^2\left(\frac{\zeta  K_2}{\sqrt{E^1}}\right)\\
&+\frac{\zeta  E^2 }{8  }\left(\frac{\partial_xE^1}{E^2}\right)^2\left(\frac{K_2}{E^1}-\frac{K^1}{E^2}\right) \sin \left(\frac{2\zeta  K_2}{\sqrt{E^1}}\right).
\end{aligned}
\end{equation}

Both $H_{\rm eff}^{(1)}$ and $H_{\rm eff}^{(2)}$ recover the classical Hamiltonian constraint in spherically symmetric GR as the quantum parameter  $\zeta$ approaches $0$. Therefore, they may serve as candidates of effective Hamiltonian constraints for certain quantum gravity theories. Since $M_{\rm eff}^{(1)}$ and $M_{\rm eff}^{(2)}$ are derived by considering the polymerization, it is expected that the underlying quantum gravity theories are related to LQG. The connection between our models and those introduced in LQG will be discussed latter.

\section{Effective metrics}
Let us take $H_{\rm eff}^{(1)}$ as an example to demonstrate how to solve the dynamics and obtain the stationary effective metrics in the corresponding model,  following the procedure in \cite{Cafaro:2024vrw}.
First, as the stationarity requires the condition  $\partial_tE^I=0$, the 
 areal gauge $E^1(x)=x^2$  can be chosen to fix the gauge generated by the diffeomorphism constraint. Second, due to the stationarity condition,  the lapse function $N$ and shift vector $N^x$ will be fixed as phase space functions by the stationary equations $\{E^I(x),H_{\rm  eff}^{(1)}[N]+H_x[N^x]\}=0$ for $I=1,2$. Third, requiring $H_x(x)=0$, $H_{\rm eff}^{(1)}=0$ and the gauge fixing condition $N^x=0$, we can solve for $E^2$ as a specific function of $x$. Note that the stationary condition ensures the static condition $N^x=0$ by the spherical symmetry.
 Additionally, requiring $N^x=0$ implies that we choose the  Schwarzschild-like coordinate to express our metric. 
 Finally, plugging the expressions for $E^I$ into Eq.~\eqref{eq:g4eff}, we get the 
 expression $\dd s_{(1)}^2$ for the metric. It  should be noted that $H_x=0$ and $H_{\rm eff}^{(1)}=0$ in the third step lead to a constant $M_{\rm eff}^{(1)}\equiv M$.  Following  the same procedure, we can also obtain the metric $\dd s_{(2)}^2$ in the theory corresponding to  $H_{\rm eff}^{(2)}$. The two effective metrics read respectively
\begin{equation}\label{eq:ds1}
\begin{aligned}
    \dd s^2_{(1)}=&-f_1\dd t^2+f_1^{-1}\dd x^2+x^2\dd\Omega^2,\\
    f_{1}=&1-\frac{2M}{ x}+\frac{\zeta^2}{x^2}\left(1-\frac{2M}{x}\right)^2,
\end{aligned}
\end{equation}
and
\begin{equation}\label{eq:metric2}
\begin{aligned}
    \dd s^2_{(2)}=&-f_2\dd t^2+\mu_2^{-1}f_2^{-1}\dd x^2+x^2\dd\Omega^2,\\
    f_{2}=&1-\frac{2M}{x},\ \mu_{2}=1+\frac{\zeta^2}{x^2}\left(1-\frac{2M}{x}\right).
\end{aligned}
\end{equation}

\subsection{Spacetime structure of $\dd s_{(1)}^2$}
The function $f_{1}$ in Eq.~ \eqref{eq:ds1} has two positive roots for all $M>0$: $x_+=2M$ and $x_-=\zeta^2/\beta-\beta/3,$ with $\beta^3=3\zeta^2  \left(\sqrt{81 M^2+3\zeta^2 }-9 M\right).$ This feature indicates that the spacetime described by $\dd s_{(1)}^2$ exhibits a double-horizon structure, similar to the normal Reissner-Nordström BH \cite{Hawking_Ellis_1973}, where the outer and inner horizons locates at $x_+$ and $x_-$ respectively. While the classical Schwarzschild singularity is resolved in this spacetime by a transition region connecting a BH to a white hole, a time-like singularity persists at $x = 0$.
Actually, such a causal structure has been observed in loop quantum BH model from various approaches (see, e.g., \cite{Modesto:2008im,Munch:2021oqn,Lewandowski:2022zce}).

If we define the effective energy-momentum tensor by $T^{\rm q}_{\rho\sigma}=G_{\rho\sigma}/(8\pi )$, with $G_{\rho\sigma}$ being the Einstein tensor of the quantum modified metric $\dd s_{(1)}^2$, a Killing observer will perceive effective energy density
\begin{equation}
\rho^{\mathrm{q}}=\frac{\zeta ^2 (x-6 M) (x-2 M)}{8\pi x^6}.
\end{equation}
For $x \gg M$, $\rho^{\rm q}$ is approximately $\zeta^2/(8\pi x^4)$. This effective density might provide a portion of the dark matter. Due to $\zeta^2\sim \hbar$, this effect is purely  quantum.

\subsection{Spacetime structure of $\dd s_{(2)}^2$}
The causal structure of the spacetime described by $\dd s_{(2)}^2$ is shown in Fig. \ref{fig:pends2} (see Appendix \ref{app:B1} for the procedure to derive the diagram). The entire spacetime comprises of three types of regions: the asymptototic flat region $A$, the BH region $B$ and the white hole region $W$. The classical singularity is replaced by a transition surface $\mathcal T$ connecting the region $B$ and region $W$. Similar diagrams are also  found in other loop quantum BH models \cite{Ashtekar:2018lag,AlonsoBardaji:2023bww}. 

The regions $A\cup B$ and $W\cup A$ in Fig. \ref{fig:pends2} can be covered by the  Painlev\'e-Gullstrand-like coordinates in which the metric is expressed as
\begin{equation}\label{eq:PGcoor}
\dd s^2_{(2)}=-\dd t^2+\frac{1}{\mu_{2}}\left(\dd x\pm \sqrt{\mu_{2}(1-f_{2})}\dd t\right)^2+x^2\dd\Omega^2.
\end{equation}
To cover the entire homogeneous region $B\cup W$, we can introduce a new coordinate $(T,X)$ with $0<T<\pi$ to rewrite the metric \eqref{eq:metric2} as
\begin{equation}\label{eq:ds2hom}
\dd s^2_{(2)}=-N^2\dd T^2+\left(\frac{2M}{R}-1\right)\dd X^2+R^2\dd\Omega^2,
\end{equation}
where $R$ and $N$, as functions of $T$, are given by 
\begin{equation*}
R+\frac{R^3}{\zeta^2}\sin^2(T)=2M,
\ N=\frac{2\zeta R^2}{\zeta^2+3R^2\sin^2(T)}.
\end{equation*}
Note that this equation implies $x = R(T) > x_0$. Hence 
the region described by the coordinate system $(t,x)$ in Eq.~\eqref{eq:metric2} with $x<x_0$ does not appear in the maximally extended spacetime. Consequently, the maximally extended spacetime is singularity free. 

With the coordinates $(T,X)$, it can be verified that the transition surface $\mathcal T$ occurs at $T=\pi/2$ and is spacelike. Moreover, the expansions $\theta_\pm(T)$ of the two null normals to the 2-spheres in the region $B\cup W$ can be calculated. As expected, $\theta_\pm(T)$ are both negative in region $B$, positive in the region $W$, and vanishing at $\mathcal T$, indicating that the black-white hole transition occurs at $\mathcal T$. Furthermore, the Kretschmann scalar $\mathcal K$ is bounded throughout the entire spacetime and reaches its maximum value at $\mathcal T$:
\begin{equation}\label{eq:KR}
\begin{aligned}
\mathcal K\big|_{\mathcal T}&=\frac{81}{4\zeta^4}+O((M\zeta^5)^{-2/3}),\\
%R\big|_{\mathcal T}&=\frac{9}{2\zeta^2}+O((M\zeta^2)^{-2/3}),
\end{aligned}
\end{equation}
implying that the transition surface $\mathcal T$ lies within the purely quantum region. 

We conclude this section by comparing the singularity natures of the maximal extensions of $\dd s_{(1)}^2$ and $\dd s_{(2)}^2$. 
In the Schwarzschild-like coordinates of $\dd s_{(1)}^2$, the domain $x>0$ is divided into three regions: $x>x_+$, $x_-<x<x_+$ and $0<x<x_-$. 
The spacetime of $\dd s_{(1)}^2$ can be extended beyond  $x=x_+$  by introducing Painlevé-Gullstrand-like coordinates  $(\tilde{t}, x)$, where the metric takes the form
\begin{equation}\label{eq:ds1PG}
\dd s^2_{(1)}=-\dd \tilde t^2+\left(\dd x\pm \sqrt{1-f_{1}}\dd \tilde t\right)^2+x^2\dd\Omega^2.
\end{equation}
Due to the square root term, the spacetime is truncated at $x=x_b$ where $x_b$ denotes the root of $1-f_1$. It turns out that $x_b<x_-$, implying that the Painlevé-Gullstrand-like coordinates already cover part of the third region  $0 < x < x_-$. By switching to Schwarzschild-like coordinates with  $x < x_-$, the maximal extension of $\dd s_{(1)}^2$ naturally reaches the central curvature singularity. In comparison with the metrics \eqref{eq:ds1PG}, the metric \eqref{eq:PGcoor} contains the factor  $1/\mu_2$, which introduces a coordinate singularity in the Painlevé-Gullstrand-like at  $x = x_0$ which is the real root of $\mu_2$. Since this singularity is also the coordinate singularity of the Schwarzschild-like coordinates, we have to introduce a new coordinate rather than switch back to the Schwarzschild-like coordinates for the extension. Consequently, the region $0<x<x_0$ cannot be included in the maximal extension of $\dd s_{(2)}^2$, leading to its free of singularity. 

\begin{figure}
    \centering
    \includegraphics[width=0.45\textwidth]{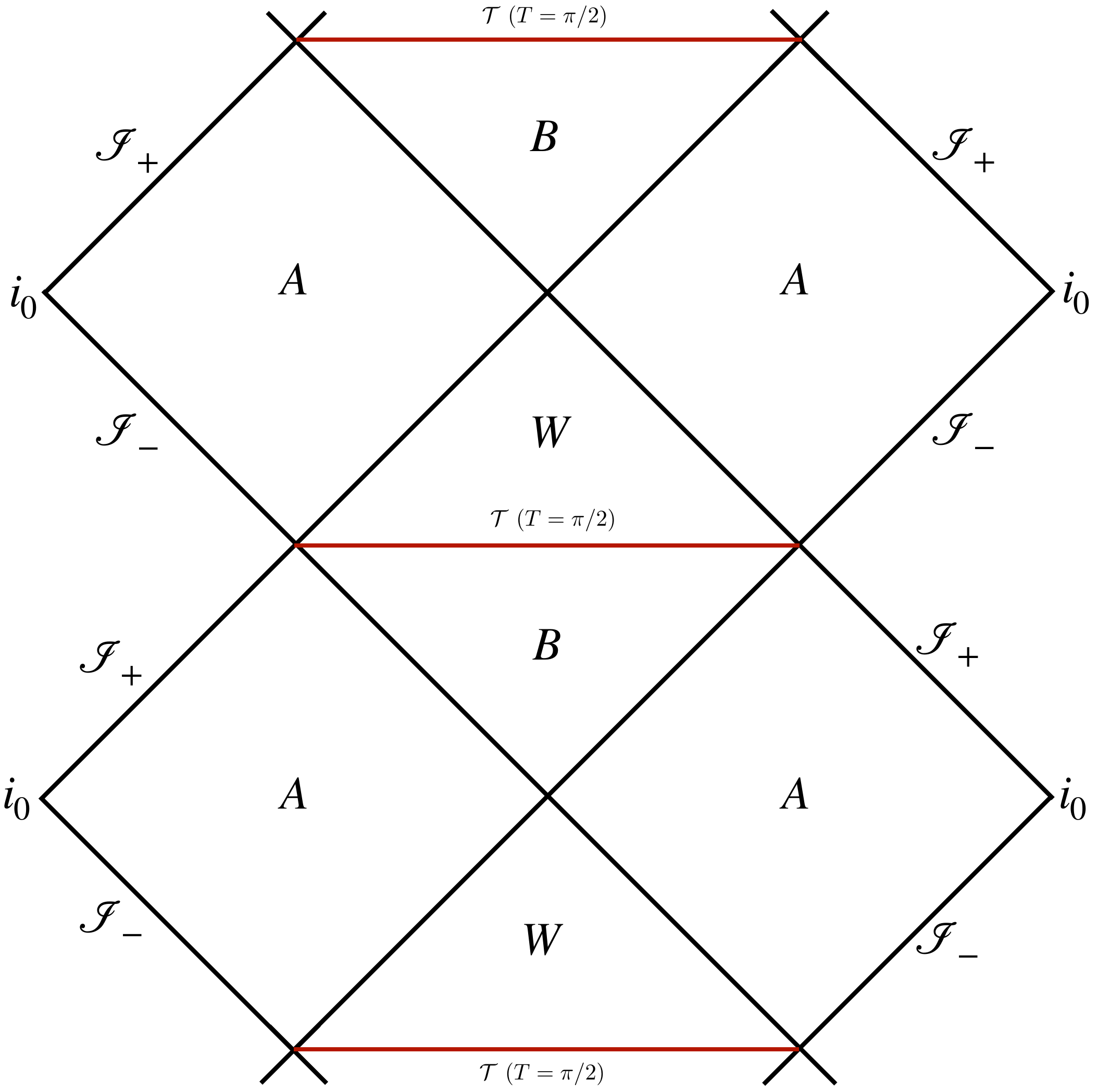}
    \caption{The Penrose diagram of the maximally extended spacetime described by $\dd s_{(2)}^2$. }\label{fig:pends2}
\end{figure}

\section{Comparison and outlook}
In loop quantum BH models such as those in \cite{Husain:2021ojz,Husain:2022gwp,Giesel:2023hys,Giesel:2023tsj}, the effective Hamiltonian was obtained by applying the standard polymerization procedure directly on the classical Hamiltonian constraint. The resulting effective Hamiltonian constraint can be alternatively obtained by substituting $\tilde M_{\rm eff}=\sqrt{s_1}/2+\sqrt{s_1}^3\sin^2\left(\zeta s_2/\sqrt{s_1}\right)/(2\zeta^2)-\sqrt{s_1}(s_4)^2/8$
for $M_{\rm eff}$ in Eq.~\eqref{eq:HeffOnMeff}. However, $\tilde M_{\rm eff}$ is not a solution to Eq.~\eqref{eq:covarianceMeff}. This explains the observed non-covariance of those models in our sense. 
Interestingly, even though the effective Hamiltonian used in those works is not equal to $H_{\rm eff}^{(1)}$, the metric $\dd s_{(1)}^2$ derived from $H_{\rm eff}^{(1)}$ is one of the solutions to the dynamics in those models with certain gauge choice \cite{Cafaro:2024vrw}. This suggests a connection between
our covariant model corresponding to $H_{\rm eff}^{(1)}$ and the loop quantum BH models.

In other loop quantum BH models \cite{Alonso-Bardaji:2021yls,Alonso-Bardaji:2022ear,AlonsoBardaji:2023bww}, a canonical transformation labeled by a constant $\lambda$ is introduced alongside the standard polymerization procedure to ensure covariance. The effective Hamiltonian in these models is similar to $H_{\rm eff}^{(2)}$, leading to the spacetime structure resembling Fig. \ref{fig:pends2}. However, an issue arises as the transition surface in these models for large BHs appears in a classical region, which is physically unacceptable. In contrast, in our spacetime $\dd s_{(2)}^2$, the transition surface for large BHs consistently lies within the Planck regime, as shown in Eq.~\eqref{eq:KR}.

In \cite{Alonso-Bardaji:2023vtl,Bojowald:2023djr,Bojowald:2023vvo,Belfaqih:2024vfk}, general covariance is analyzed via a similar approach to ours. The resulting effective Hamiltonians resemble $H_{\rm eff}^{(2)}$, but with unspecified free functions and a fixed form for $\mathcal{R}$. These free functions emerge as integration constants in Eq.~\eqref{eq:covarianceMeff} in this work. However, covariant constraints akin to $H_{\rm eff}^{(1)}$ are not considered, and the polynomial ansatz for the Hamiltonian constraint limits these models from capturing other covariant Hamiltonians beyond this structure (see \cite{Zhang:2024ney} for an example).

The approach developed in this letter to study covariant effective models of quantum gravity can be extended to include matter coupling (see. \cite{Zhang:2024ney} for more details).
Note that covariance does not uniquely determine a model but constrains the available models. For instance, the models in \cite{Husain:2021ojz,Husain:2022gwp,Giesel:2023hys,Giesel:2023tsj,Ashtekar:2018lag,Zhang:2021xoa} fail to meet the conditions \eqref{eq:HeffOnMeff} and \eqref{eq:covarianceMeff}. This work also opens avenues for further research, such as connecting effective Hamiltonians \eqref{eq:heff1} and \eqref{eq:heff2} to full LQG, and extending beyond spherical symmetry.

\begin{acknowledgments}
%C.Z. acknowledges the valuable discussions with Asier Alonso Bardaj\'\i, Martin Bojowald, David Brizuela, Luca Cafaro, Kristina Giesel, Hongguang Liu, Chun-Yen Lin, Dongxue Qu, Farshid Soltani, Edward Wilson-Ewing, and Stefan Andreas Weigl. 
CZ thanks to Kristina Giesel and Hongguang Liu for their proposal concerning the constraint algebra at the early stage of the work. This  work is supported by the National Science Centre, Poland as part of the OPUS 24 Grant No. 2022/47/B/ST2/02735, NSFC Grant No. 12275022 and NSFC Grant No. 12165005.

\end{acknowledgments}

\bibliography{reference}

\appendix
\include{supplemental_material_v7_final}

\end{document}

%% file: supplemental_material_v7_final.tex
\onecolumngrid

\section{Conditions of covariance}\label{app:A}
The dynamics in the Hamiltonian formulation of the spherically symmetric model can be understood in two perspectives.   On one hand,  solution to the Hamilton's equation with $\heff[N] + H_x[N^x]$ as the Hamiltonian yields a curve $t \mapsto (K_I^{(t)}(x), E^I_{(t)}(x))$ in the phase space. On the other hand, the quantities $K_I(t,x) \equiv K_I^{(t)}(x)$, $E^I(t,x) \equiv E^I_{(t)}(x)$, as well as $N(t,x) \equiv N^{(t)}(x)$ and $N^x(t,x) \equiv N^x_{(t)}(x)$, can be interpreted as fields on $\mathcal{M}_2$, when the fields at each time $t$ are pushed forward from $\Sigma$ to $\Sigma_t$. Combining the two perspectives, we can rewrite the Hamilton's equations in terms of the Lie derivatives on $\mathcal M_2$ as  
\begin{equation}\label{eq:effHamton2}
\begin{aligned}
\mathcal L_{\mathfrak N}E^1&=\{E^1,\heff[N]\},\\
\mathcal L_{\mathfrak N}E^2&=\{E^2,\heff[N]\},\\
\mathcal L_{\mathfrak N}K_1&=\{K_1,\heff[N]\},\\
\mathcal L_{\mathfrak N}K_2&=\{K_2,\heff[N]\},
\end{aligned}
\end{equation} 
where $E^1$ and $K_2$ are scalar fields, and $E^2$ and $K_1$ are scalar densities of weight $1$. 

Then, we consider an infinitesimal gauge transformation on the phase space generated by $\heff[\alpha N]$. This transformation maps the curve $t\mapsto (K_I^{(t)}(x), E^I_{(t)}(x))$ in the phase space to the new curve $t\mapsto (K_I^{(t)}(x) +\epsilon \delta K_I^{(t)}(x), E^I_{(t)}(x) +\epsilon \delta E^I_{(t)}(x))$, where
\begin{equation*}
\begin{aligned}
\delta K_I(x)=&\{K_I(x),\heff[\alpha N]\},\\
\delta E^I(x)=&\{E^I(x),\heff[\alpha N]\}.
\end{aligned}
\end{equation*}
 By requiring that the new curve satisfies Hamilton’s equation \eqref{eq:effHamton2} with respect to the new lapse function, $N + \epsilon \delta N$, and the new shift vector, $N^x + \epsilon \delta N^x$, we derive $\delta N$ and $\delta N^x$ as
\begin{equation}\label{eq:dNN}
\begin{aligned}
\delta N^x\approx &-N^2\mu E^1(E^2)^{-2}\partial_x\alpha,\\
\delta N\approx &\mathcal L_{\alpha \mathfrak N }N+N\mathfrak N^\rho \partial_\rho \alpha.
\end{aligned}
\end{equation}
Here, the notation $A \approx B$ is referred to as $A$ being weakly equal to $B$, which indicates that $A$ is equal to $B$ when the constraints vanish.

Since $H_{\rm eff}$ could depend on $\partial_x^nK_I$, we have
\begin{equation}\label{eq:E1d}
\{E^I(x),\heff[\alpha N]\}\approx \alpha(x)\{E^I(x),\heff[N]\}+\Delta_I,
\end{equation}
where
\begin{equation}\label{eq:defineD1}
\begin{aligned}
\Delta_I=&-\sum_{n\geq 1}(-1)^n\left[\sum_{m=1}^n\binom{n}{m}(\partial_x^m\alpha)\frac{\partial^n}{\partial x^{n-m}}\left( N(x)\frac{\partial \heff(x)}{\partial(\partial_x^nK_I(x))}\right)\right].
\end{aligned}
\end{equation}
Because $E^1$ is a scalar, leading to $\alpha \mathcal L_{\mathfrak N}E^1=\mathcal L_{\alpha \mathfrak N}E^1$, Eq.~\eqref{eq:E1d} for $I=1$ can be simplified as
\begin{equation}\label{eq:dE1}
\delta E^1\approx \mathcal L_{\alpha \mathfrak N}E^1+\Delta_1.
\end{equation}
Similarly, since $E^2$ is a scalar density of weight $1$, we get 
\begin{equation}\label{eq:dE2}
\delta E^2\approx \mathcal L_{\alpha \mathfrak N }E^2-E^2\mathfrak N^\rho \partial_\rho\alpha+\Delta_2.
\end{equation}
Using Eqs.~\eqref{eq:dNN}, \eqref{eq:dE1} and  \eqref{eq:dE2}, we  finally get
\begin{equation}\label{eq:s7p}
\begin{aligned}
&\delta g^{(\mu)}_{\rho\sigma}\dd x^\rho\dd x^\sigma\approx\mathcal L_{\alpha\mathfrak N}(g_{\rho\sigma}^{(\mu)}\dd x^\rho\dd x^\sigma)+\Delta_1\dd\Omega^2-\left[\delta(\mu E^1)-\mathcal L_{\alpha\mathfrak N}(\mu E^1)-\frac{2\mu}{E^2}\Delta_2\right]\frac{(\dd x+N^x\dd t)^2}{\mu E^1S}.
\end{aligned}
\end{equation}
Note that, in Eq.~\eqref{eq:s7p}, if $\alpha$, $N$ and $N^x$ are phase space independent, the weak equality becomes the strong equality.
In addition, it can be verified that,  once Eq.~\eqref{eq:s7p} is strongly satisfied for all phase space independent quantities $\alpha$, $N$ and $N^x$,  it will weakly hold for all phase space dependent $\alpha$, $N$ and $N^x$. Therefore, it is necessary and sufficient to consider Eq.~\eqref{eq:s7p} for all phase space independent quantities $\alpha$, $N$ and $N^x$ in what follows. 

According to Eq.~\eqref{eq:s7p}, to ensure $\delta g^{(\mu)}_{\rho\sigma}=\mathcal L_{\alpha\mathfrak N}g^{(\mu)}_{\rho\sigma}$, it is sufficient and necessary that the following two equations are satisfied for all phase-space-independent $\alpha$, $N$ and $N^x$:
\begin{subequations}
\begin{align}
\Delta_1=&0,\label{eq:condition1}\\
\delta(\mu E^1)=&\mathcal L_{\alpha\mathfrak N}(\mu E^1)+\frac{2\mu}{E^2}\Delta_2\label{eq:condition2}.
\end{align}
\end{subequations}
By Eq.~\eqref{eq:defineD1}, the first condition \eqref{eq:condition1} leads to
\begin{equation}
\frac{\partial \heff(x)}{\partial(\partial_x^nK_1(x))}=0,
\end{equation}
which implies that $\heff$ is independent of the derivatives of $K_1$. Additionally, due to Eq.~\eqref{eq:E1d} with $I=2$, Eq.~\eqref{eq:condition2} is equivalent to 
\begin{equation}\label{eq:dmulmu}
\alpha(x)\left\{S(x),\heff[N]\right\}=\left\{S(x),\heff[\alpha N]\right\},
\end{equation}
for all phase space independent $\alpha$ and $N$. It should be noted that the use of the strong equality rather than the weak equality in Eq.~\eqref{eq:dmulmu} ensures that the resulting expression of $\mu$ remains valid across various matter coupling models.

\section{Construct Penrose diagram of $\dd s^2_{(2)}$}\label{app:B1}
We now analyze the causal structure of the spacetime defined by $\dd s_{(2)}^2$. In the Schwarzschild-like coordinates, the line element has two coordinate singularities located at  $x = 2M$, corresponding to the root of  $f_2$, and at  $x = x_0\equiv \zeta^2/\beta-\beta/3$ with $\beta^3=3\zeta^2  \left(\sqrt{81 M^2+3\zeta^2 }-9 M\right)$, corresponding to the real root of $\mu_2$. Thus,  the domain $x>0$ is divided into three separate regions by the two coordinate singularities, and the maximal extension of one of these regions does not necessarily include the others. We begin with the region $x > 2M$. To extend this region to include the boundary  $x = 2M$,  we adopt the Painlev\'e-Gullstrand-like coordinates  $(\tilde t, x)$, defined by the transformation
\begin{equation}
\dd \tilde t=\dd t\pm \frac{\sqrt{\mu_2(1-f_2)}}{\mu_2f_2} \dd x.
\end{equation}
In the new coordinates, the metric becomes
\begin{equation}\label{eq:PGcoor}
\dd s^2_{(2)}=-\dd \tilde t^2+\frac{1}{\mu_{2}}\left(\dd x\pm \sqrt{\mu_{2}(1-f_{2})}\dd \tilde t\right)^2+x^2\dd\Omega^2.
\end{equation}
It is straightforward to show that $\dd s^2_{(2)}$ in Eq.~\eqref{eq:PGcoor} is still a solution to the theory governed by  $H_{\rm eff}^{(2)}$, by following the previous procedure but with an alternative gauge-fixing condition  $E^2 = x$  applied in the third step to solve for  $N^x$.

To further extend the spacetime beyond the coordinate singularity $x=x_0$ in the Painlev\'e-Gullstrand-like coordinates, we adopt the new coordinates $(T, X)$ through the following transformations 
\begin{equation}
x = R,\quad \dd \tilde t = \dd X \pm \frac{\sqrt{\mu_2(1-f_2)} (\partial_TR )}{\mu_2f_2}\dd T
\end{equation}
where  $R$  is a function of  $T$  given by
\begin{equation}\label{eq:defineR}
R+\frac{R^3}{\zeta^2}\sin^2(T)=2M.
\end{equation}
In these new coordinates, the metric takes the form
\begin{equation}\label{eq:ds2hom}
\dd s^2_{(2)}=-N^2\dd T^2+\left(\frac{2M}{R}-1\right)\dd X^2+R^2\dd\Omega^2,
\end{equation}
where the range of the coordinates reads $0<T<\pi$ and $X\in\mathbb R$, and $N$ is given by
\begin{equation}
N=\frac{2\zeta R^2}{\zeta^2+3R^2\sin^2(T)}.
\end{equation}
This new coordinate system is homogeneous in space since the metric is independent of  $X$. Eqs. \eqref{eq:defineR} and \eqref{eq:ds2hom} indicate that there is neither coordinate singularity nor curvature singularity in the domain of $(T, X)$. Further details regarding the verification that $\dd s^2_{(2)}$ in Eq.~\eqref{eq:ds2hom} satisfies the dynamics of  $H_{\rm eff}^{(2)}$  can be found in App. \ref{app:B}. The causal structure of the spacetime described by $\dd s_{(2)}^2$ can be obtained by gluing the charts of the coordinate systems $(\tilde t, x)$ and $(T,X)$, as
is shown in Fig. 1.

\section{Homogeneous solution corresponding to $H_{\rm eff}^{(2)}$}\label{app:B}

Let us consider the homogeneous solution where we require
\begin{equation}
\partial_xE^I=0=\partial_xK_I.
\end{equation}
In this gauge, the diffeomorphism constraint is solved automatically, and  the Hamiltonian constraint is reduced to
\begin{equation}
\begin{aligned}
H_{\rm eff}^{(2)}=&-\frac{3 \sqrt{E^1}E^2}{2 \zeta ^2 } \sin ^2\left(\frac{\zeta  K_2}{\sqrt{E^1}}\right)+\frac{K_2E^2 }{2\zeta  } \sin \left(\frac{2\zeta  K_2}{\sqrt{E^1}}\right)-\frac{ E^1 K_1}{2\zeta  }\sin \left(\frac{2\zeta  K_2}{\sqrt{E^1}}\right)-\frac{E^2}{2 \sqrt{E^1}}.
\end{aligned}
\end{equation}
The constant of motion $M_{\rm eff}^{(2)}$ becomes
\begin{equation}\label{eq:mexpress}
M_{\rm eff}^{(2)}=\frac{\sqrt{E^1}}{2}+\frac{\sqrt{E^1}^3}{2\zeta^2}\sin^2\left(\frac{\zeta K_2}{\sqrt{E^1}}\right).
\end{equation}

The homogeneous solution is obtained by choosing 
\begin{equation}
\frac{\zeta K_2}{\sqrt{E^1}}=t,
\end{equation}
where $t$ represents the time coordinate. Then, Eq. \eqref{eq:mexpress} can be simplified as
\begin{equation}\label{eq:Msint}
M_{\rm eff}^{(2)}=\frac{\sqrt{E^1}}{2}+\frac{\sqrt{E^1}^3}{2\zeta^2}\sin^2(t)=M,
\end{equation}
leading to
\begin{equation}
\sqrt{E^1(t)}\equiv R(t)=\frac{1}{\sin(t)}\left(\frac{\zeta^2}{\beta(t)}-\frac{\beta(t)}{3}\right),
\end{equation}
where 
\begin{equation}
 \beta(t)^3=3\zeta^2\left(-9M\sin(t)+\sqrt{3\zeta^2+81M^2\sin^2(t)}\right).
\end{equation}
Moreover, we have 
\begin{equation}\label{eq:dotR}
\dot R(t)=\{\sqrt{E^1},\heff[N]\}=\frac{N}{2\zeta}R(t)\sin(2 t).
\end{equation}
To get $N$, let us consider 
\begin{equation}
\begin{aligned}
0=&\dot M_{\rm eff}^{(2)}=-\frac{\zeta ^2 \dot R(t)+3 R(t)^2 \sin ^2(t) \dot R(t)+R(t)^3 \sin (2 t)}{\zeta ^2 }.
\end{aligned}
\end{equation}
Substituting Eq.~\eqref{eq:dotR}, we get
\begin{equation}\label{eq:NNR}
\begin{aligned}
N=&-\frac{2 \zeta  R(t)^2}{\zeta ^2+3 R(t)^2 \sin ^2(t)}.
\end{aligned}
\end{equation}
Furthermore, we have
\begin{equation}\label{eq:dotE2}
\begin{aligned}
\dot E^2(t)=\frac{N [E^2 (\sin (2 t)-t \cos (2 t))+\zeta  K_1(t) R(t) \cos (2 t)]}{\zeta }.
\end{aligned}
\end{equation}
The constraint equation $H_{\rm eff}^{(2)}=0$ results in
\begin{equation}\label{eq:K1}
\begin{aligned}
K_1=& -\frac{E^2}{\zeta  (E^1)^{3/2}} \Bigg[\zeta ^2 \csc \left(\frac{2 \zeta  K_2}{\sqrt{E^1}}\right)-\zeta  \sqrt{E^1} K_2+\frac{3}{2} E^1 \tan\left(\frac{\zeta  K_2}{\sqrt{E^1}}\right) \Bigg].
\end{aligned}
\end{equation}
According to Eq.~\eqref{eq:dotR}, we obtain
\begin{equation}\label{eq:Npppp}
N(t)=\frac{2\zeta \dot R(t)}{R(t)\sin(2t)}.
\end{equation}
Combining Eqs.~\eqref{eq:dotE2}, \eqref{eq:K1} and \eqref{eq:Npppp}, we get
\begin{equation}\label{eq:dotE2p}
\begin{aligned}
\frac{\dot E^2(t)}{E^2(t)}=&\frac{2\zeta \dot R(t)}{R(t)\sin(2t)}\Bigg(-\frac{\zeta  \cot (2 t)}{R(t)^2}-\frac{ (\sin (3 t)-5 \sin (t)) \sec (t)}{4 \zeta }\Bigg).
\end{aligned}
\end{equation}
Equation \eqref{eq:Msint} leads to
\begin{equation}
\sin^2(t)=\frac{2M\zeta^2}{R(t)^3}-\frac{\zeta^2}{R(t)^2}.
\end{equation}
Substituting this result into Eq.~\eqref{eq:dotE2p}, we finally get
\begin{equation}
E^2(t)=\frac{\sqrt{2 M-R(t)} \sqrt{-2 \zeta ^2  M+\zeta ^2 R(t)+R(t)^3}}{R(t)}.
\end{equation}
Furthermore, we have
\begin{equation}
\mu_2(t)=\frac{-2M\zeta^2+\zeta^2 R(t)+R(t)^3}{R(t)^3}.
\end{equation}
Combining all of the above results,  we get the metric
\begin{equation}\label{eq:metrichom}
\begin{aligned}
\dd s^2_{(2)}=&-\frac{4\zeta^2R(t)^4}{\left(\zeta^2+3R(t)^2\sin^2(t)\right)^2}\dd t^2+\left(\frac{2M}{R(t)}-1\right)\dd x^2+R(t)^2\dd\Omega^2. 
\end{aligned}
\end{equation}
In the main text, $(T,X)$ instead of $(t,x)$ is used for clarity.

%To see that the metric \eqref{eq:metrichom} is the same as the one under the PG coordinate, let us use $T,\ X$ instead of $t,\ x$ to rewrite the metric \eqref{eq:metrichom} as
%\begin{equation}\label{eq:metrichom2}
%\begin{aligned}
%\dd s^2_{(2)}=&-\frac{4\zeta^2R(T)^4}{\left(\zeta^2+3R(T)^2\sin^2(T)\right)^2}\dd T^2\\
%&+\left(\frac{2GM}{R(T)}-1\right)\dd X^2+R(T)^2\dd\Omega^2. 
%\end{aligned}
%\end{equation}
%Then, the PG coordinate $(\tilde t,x)$ is related to  $(T,X)$ by
%\begin{equation}
%x=R(T),\quad \dd \tilde t=\dd X+\alpha \dd T
%\end{equation}
%with 
%\begin{equation}
%\alpha=\pm \mu_2^{-1}\dot R(T)\sqrt{\mu_2\frac{2M}{R(T)}}\left(1- \frac{2M}{R(T)}\right)^{-1}.
%\end{equation}
%

%% file: covariant_LQBH_paperversionArxiV3.bbl
\begin{thebibliography}{53}
\bibitem [{not()}]{note2}%
  \BibitemOpen
  \href@noop {} {}\bibinfo {note} {Dirac observables are those functions of the
  canonical variables that Poisson commute with the first-class constraints
  generating gauge transformations.}\BibitemShut {Stop}%
\end{thebibliography}
